\documentclass[twocolumn,showpacs,pra]{revtex4}

\usepackage{graphicx}
\usepackage{rotating}
\usepackage{amsmath}
\usepackage{amsfonts}
\usepackage{amssymb}
\usepackage{enumerate}
\usepackage{longtable}
\usepackage{dcolumn}
\usepackage{bbm}
\usepackage{subfigure}
\usepackage{color}

\begin{document}

\newcommand{\bb}{_\mathrm{b}}
\newcommand{\pp}{_\mathrm{p}}
\newcommand{\sss}{_\mathrm{s}}
\newcommand{\rudd}{^\mathrm{RUDD}}
\newcommand{\udd}{^\mathrm{UDD}}
\newcommand{\red}[1]{\textcolor{red}{#1}}

\title{Symmetry-Enhanced Performance of Dynamical Decoupling}

\author{S. Pasini}
\email{s.pasini@fz-juelich.de}
\affiliation{Forschungszentrum J\"ulich, 52425 J\"ulich, Germany}

\author{G. S. Uhrig}
\email{goetz.uhrig@tu-dortmund.de}
\affiliation{Lehrstuhl f\"{u}r Theoretische Physik I, 
Technische Universit\"{a}t Dortmund,
 Otto-Hahn Stra\ss{}e 4, 44221 Dortmund, Germany}
\date{\rm\today}

\begin{abstract}
We consider a system with general decoherence  and a quadratic dynamical 
decoupling sequence (QDD) for the coherence control of a qubit coupled to a 
bath of spins. 
We investigate the influence of the geometry and of the initial conditions of 
the bath on the performance of the sequence. The overall performance is 
quantified by a distance norm $d$. It is expected that $d$ scales with $\tau$, 
the total duration of the sequence, as $\tau^{\min \{N_x,N_z\}+1}$, where $N_x$ and 
$N_z$ are the number of pulses of the outer and of the inner sequence, 
respectively. We show both numerically and analytically that the state of the
bath can boost the performance of QDD under certain conditions: 
The scaling of QDD 
for a given number of pulses can be enhanced by a factor of 2 if the bath is 
prepared in a highly symmetric state and if the system Hamiltonian is SU(2) 
invariant. 
\end{abstract}


%

\maketitle

\section{Introduction}

Improvements both in resonance spectroscopy and in quantum information rely on 
the ability of suppressing unwanted couplings between the system and its 
environment. Uncontrolled couplings  are often the origin of phase accumulation
 and in general of decoherence. Therefore, a faithful manipulation and 
preservation of quantum states is required.

The dynamical decoupling (DD) is an open-loop control scheme to average out the
 undesired coupling between the system (qubit) and the environment (bath) by 
means of stroboscopic pulsing of the qubit. The DD was developed by Viola and 
Lloyd \cite{viola98} from the original idea of Hahn \cite{hahn50}. 

In its original formulation the DD makes use of equidistant $\pi$ pulses to 
average out only a single coupling along one spin direction, usually the $z$
direction, (pure dephasing) - we think for example of 
the Carr, Purcell, Maiboom and Gill (CPMG) sequence \cite{carr54,meibo58}. A 
remarkable advance is the optimal DD discovered by Uhrig \cite{uhrig07}, whose 
sequence has the minimun number of pulses for a given order of the suppression 
of the decoherence. It was shown that UDD can also be used to suppress 
longitudinal relaxation \cite{yang08,uhrig08,uhrig09b}. Recently other 
non-equidistant sequences have been proposed \cite{bierc09a,uys09,khodj11a}.

The most general case concerns the suppression of dephasing and longitudinal
relaxation at 
the same time. A sequence of pulses having a single level of suppression 
cannot suppress general dephasing. Sequences with two sorts of pulses have 
been proposed where concatenated sequences are used, like CDD \cite{khodj05} 
and the CUDD \cite{uhrig09b} for example.

Recently West \textit{et al.} \cite{west10} have proposed a near optimal 
scheme that suppresses arbitrary couplings to order $\tau^N$ ($\tau$ is the duration 
of the total sequence) between the qubit and the bath using ${\cal O}(N^2)$ 
pulses. The sequence consists of two levels of nested UDD, therefore the name 
quadratic UDD (QDD). 
The validity of UDD can be extended to analytically time-dependent 
Hamiltonians \cite{pasin08b} which is an important ingredient for the 
demonstration of QDD. 

Wang \textit{et al.} \cite{wang11} showed that the effect of QDD can be 
decomposed in the effects of the inner and the outer sequences. The concept 
of mutually orthogonal operation set (MOOS) for nested Uhrig DD was introduced:
 a set of control operators on the inner level is not affected by a set of 
control operators on the outer level if both sets come from a MOOS. Higher 
order protection of a MOOS can be achieved if even-order UDD sequence on 
different levels are nested. Thus the results in Ref.\ \onlinecite{wang11} 
demonstrate the validity of QDD with even-order UDD sequence on the inner 
level. If the inner level has an odd order the symmetry group generated by 
MOOS is broken and the scheme based on nested UDD can not be applied anymore. 
It appears that this problem has been solved by Jiang and Imambekov 
\cite{jiang11} who have provided a proof of the validity of nested UDD (NUDD) 
sequences that relies on a mapping between NUDD and a discrete quantum walk in 
$2m$ dimensional space. The case of QDD corresponds to $m=1$. At last, an 
alternative proof of the validity of QDD and a numerical investigation of the 
scaling of the errors along specific
spin directions for QDD  has been presented in Ref.\ \onlinecite{kuo11} 
and in Ref.\ \onlinecite{quiro11}, respectively.

In this paper we want to draw the attention to the effects of the state of
the bath on the 
performance of the sequence. The fact that the specifics of the bath can limit 
the performance of a sequence is already known \cite{cywin08,pasin10a}. It was 
tested experimentally that UDD can outperform CPMG  if the environment is 
characterized by a hard cutoff \cite{bierc09a,du09} while for soft cutoffs 
equidistant sequences perform either better or the same \cite{ajoy11}. 
Otherwise UDD seems to perform very well for electron spins in irradiated 
malonic acid crystals \cite{du09}
as well as for applications of magnetic resonance imaging \cite{jenis09}. 

So far we have always viewed the environment as an unavoidable restraint on 
the prolongation of the coherence of a spin (qubit).
Hence one has to eliminate or at least to reduce the coupling between 
environment and system because the coupling between environment and system
transfers disorder from the environment to the system. 
But does the environment's disorder always act against 
coherence in the system?

Here we show that the performance of a given sequence can be enhanced if the 
system Hamiltonian $H$ is SU(2) invariant and if the initial state of the bath,
 i.e., its density matrix
$\rho_\text{B}$  is completely disordered: $\rho_\text{B}\propto \mathbbm{1}$. 
We call such a state an infinite-temperature state. We simulate the effect of 
a QDD sequence on a bath of spins both for a completely anisotropic as well as 
for an isotropic central Heisenberg spin model. For both cases we analyze the 
scaling of QDD when the bath is prepared either in a product state or in an 
infinite-temperature state. Four cases are studied as summarized in Tab.\ 
\ref{tab0}. 
\begin{table}[h]
\begin{center}
\begin{align}
\label{tab0}
\begin{tabular}{c||c|c}
$\rho_\text{B} \setminus H$ & low & high \\
\hline
\hline
low &  $\tau^{N_\text{min}+1}$ & $\tau^{N_\text{min}+1}$ \\
\hline
high & $\tau^\chi$ & $\tau^{2(N_\text{min}+1)}$ \\
\end{tabular}
\end{align}
\caption{The scaling of QDD with the duration of the sequence $\tau$ classified 
  according to the degree of symmetry of the Hamiltonian and of the initial 
  state of the bath. The notation $N_\text{min}$ refers to the minimum number of 
  pulses of the inner and the outer sequence.
}
\end{center}
\end{table}
The cases, where the bath state is characterized by a 
low degree of symmetry, independent of the degree of symmetry of $H$, 
provide the lower bounds for the scaling of QDD: For 
short times QDD scales always as $\tau^{N_\text{min}+1}$, where $N_\text{min}$ is the 
smallest number of pulses, either of the inner or of the outer sequence 
\cite{quiro11,kuo11}. 
In the off-diagonal case in Tab.\ \ref{tab0},
where $H$ is of low symmetry and the bath state of high symmetry,
 the scaling exponent $\chi$ depends on the number of pulses.
Otherwise, the scaling of QDD is enhanced to the power $\tau^{2(N_\text{min}+1)}$ if 
$H$ and $\rho_\text{B}$ are highly symmetric.

The paper is set up as follows: in Sect.\ \ref{sec1} and in the Sect.\ 
\ref{sec2} the numerical results for the low and for the high symmetry cases. 
In Sect.\ \ref{sec3} we provide the analytical argument for the appearance of 
the factor 2 in the high symmetry case. In Sect.\ \ref{sec4} we study the 
off-diagonal cases with mixed symmetry.
At last we draw our conclusions in Sect.\ \ref{conclusions}.

\section{Case 1: Low symmetry}
\label{sec1}
\begin{figure}[ht]
     \begin{center}
      \includegraphics[width=0.5\columnwidth,clip]{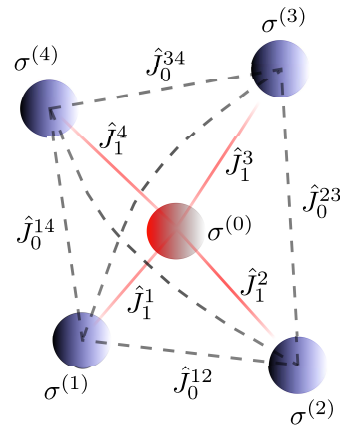} 
     \end{center}
    \caption{(Color online) Central spin model described by Eq.\ (\ref{eq:ham}) 
with $M=4$. The solid lines represent the coupling between the central spin  
    (qubit) and the spins of the bath, while the dashed lines represent the 
couplings between the spins of the bath.
    \label{fig0}
    }
\end{figure}

We start from the case where both the system Hamiltonian as well as the initial
 state of the bath have a low degree of symmetry. 
We consider a central spin model 
\cite{schli03,bortz07a,bortz07b,witze08,lee08a} 
characterized by a completely anisotropic Hamiltonian of the form
\begin{subequations}
\begin{eqnarray}
H&=&H_\text{B}+H_\text{qB}
 \label{eq:ham_a}
\\
&=& \sum_{i=1}^M\sum_{j>i}^M\vec{\sigma}^{(i)}\hat{J}_0^{ij}\vec{\sigma}^{(j)}+
\sum_{i=1}^M\vec{\sigma}^{(0)}\hat{J}_1^{i}\vec{\sigma}^{(i)},
 \label{eq:ham_b}
\end{eqnarray} 
 \label{eq:ham}
\end{subequations}
where all nine entries of the $3\times 3$ $\hat{J_0^{ij}}$ and 
$\hat{J_1^{i}}$ matrices are random numbers drawn from the
interval $[-1,1]$ (see Fig. \ref{fig0}).
The system does not show any 
symmetry. The entries for the matrices $\hat{J_0^{ij}}$ and $\hat{J_1^{i}}$ are 
fixed
randomly at the beginning of the simulation and they remain the same for all 
the numerical results we present in this article.
The spin labelled with zero represents the qubit while $M$ defines the number 
of spins in the bath. The scaling appears
to be essentially independent \cite{pasin11a} 
of $M$; we considered for our simulation $M=8$. 
Calculations for $M=3$ yield the same results as far as the
exponents are concerned.

The QDD sequence is made of an outer sequence of $\pi$-pulses about
$\sigma_x$ and of an inner sequence of $\pi$-pulses about $\sigma_z$. 
The number of pulses for each sequence is $N_x$ and $N_z$, respectively, 
and the total number of pulses of the sequence is $N_x+N_z+N_xN_z$. 
The switching instants are given by
\begin{subequations}
\begin{equation}
 \label{eq:instants_x}
t^x_j=\tau\sin^2\left[\frac{j \pi}{2(N_x+1)}\right]
\end{equation}
\begin{equation}
\label{eq:instants_z}
t^z_{k,j}=t^x_{j}+(t^x_{j+1}-t^x_{j})\sin^2\left[\frac{k \pi}{2(N_z+1)}\right] 
\end{equation}
\end{subequations} for $j=\{1,..,N_x\}$ and $k=\{1,..,N_z\}$. 
We use the notation $t^x_{N_x+1}=\tau$.

We start with an initial density matrix of the total system of the form
\begin{align}
 \label{eq:rho_0}
\rho_0^{(\gamma)}=|\gamma\rangle\langle\gamma|\otimes \rho_\text{B},
\end{align} 
with $\gamma=\{x,y,z\}$. The first factor in the tensor product
refers to the Hilbert space of the qubit, the second
to the Hilbert space of the bath.
Furthermore, we introduce the notation
\begin{align}
 \label{eq:rho_0_SB}
\rho_0^{(\gamma)}:=\rho_\text{S}^{(\gamma)}\otimes\rho_\text{B}.
\end{align}
For the low symmetry case we assume that 
the bath is initially in a pure product state
\begin{equation}
 \label{eq:rho_product}
|\psi_\text{B}\rangle = \bigotimes_{i=1}^M |\gamma_i\rangle
\end{equation}
so that $\rho_\text{B}=|\psi_\text{B}\rangle\langle\psi_\text{B}|$.
For the high symmetry case we choose
\begin{equation}
\rho_\text{B}=\mathbbm{1}_\text{B}/D
\end{equation}
where $D$ is the dimension of the Hilbert space of the bath
so that $\text{Tr}_\text{B}\rho_\text{B}=1$.
 
The overall performance of the sequence is given by the norm distance 
\cite{lidar08}
\begin{subequations}
 \label{eq:pF}
\begin{eqnarray}
d^2 &:=&\frac{1}{3}\sum_{\gamma\in\{x,y,z\}} d^2_\gamma \\
d^2_\gamma &:=& \text{Tr}_\text{q}\left[\Delta^{(\gamma)}(\tau)\right]^2,
\end{eqnarray}
\end{subequations}
with
\begin{equation}
 \label{eq:delta} \Delta^{(\gamma)}(\tau):= 
\text{Tr}_\text{B}\left(U_{\text{B}}\hat{P}\rho_0^{(\gamma)}\hat{P}^\dagger 
U_{\text{B}}^\dagger-U(\tau,0)\rho_0^{(\gamma)}U(\tau,0)^\dagger\right).
\end{equation} 
The norm distance measures the distance of the real evolution to
the ideal one.
The operator $\hat{P}$ is defined by 
\begin{equation}
\label{eq:P_op}
\hat P:=\underset{N_x+1}{\underbrace{\sigma_z^{N_z}\sigma_x\sigma_z^{N_z}
\sigma_x...\sigma_z^{N_z}}}.
\end{equation}
It incorporates the effects of the pulses. 
The operator $U(\tau,0)$ represents the evolution 
operator of the system (\ref{eq:ham}) 
\begin{equation}
 \label{eq:U}U(\tau,0)={\cal T}\left\{e^{-i\int_0^\tau H(t)\text{d}t}\right\},
\end{equation}
where ${\cal T}$ stands for the time-ordering and
\begin{equation}
 \label{eq:U_B}
U_{\text{B}}:=\mathbbm{1}\otimes e^{-i \tau H_\text{B}}
\end{equation}
is the  dynamics of the isolated bath.
During the application of the sequence the sign in front of the coupling terms 
between the qubit and the bath perpendicular to the pulse direction changes 
every time a $\pi$ pulse is applied because
$\sigma_i\sigma_j=-\sigma_j\sigma_i$ for $i\neq j$.
Thus, in the toggling frame, the system Hamiltonian (\ref{eq:ham}) can be 
written as a time dependent Hamiltonian
\begin{equation}
 \label{eq:H(t)} H(t)=H_\text{B}+\sum_{i=1}^M \sum_{j,k=1}^3f_j(t)\ 
\sigma^{(0)}_j\left(J_1^{i}\right)_{jk}\sigma_k^{(i)},
\end{equation} 
where the switching functions $f_j(t)$ are a piecewise constant functions with 
values $\pm 1$. By $\sigma_k^{(i)}$ we refer to the $k$ component of the vector 
$\vec \sigma^{(i)}$. Analogously, we refer by $\left(J_1^{i}\right)_{jk}$
to the $(j,k)$ element of the matrix $\hat J_1^i$. 

Our simulations show that for $N_x\neq 0$, $N_z\neq 0$ and 
\begin{subequations}
\begin{align}
 & \begin{array}{c} \text{for}\ N_x=N_z=N \rightarrow  d\propto \tau^{N+1}+{\cal O}
\left(\tau^{N+2}\right), \end{array}\\
 & \begin{array}{c}  \text{for}\  N_x>N_z\rightarrow d\propto \tau^{N_z+1}+{\cal O}
\left(\tau^{N_z+2}\right), \end{array}\\
 & \begin{array}{c}  \text{for}\  N_x<N_z\rightarrow d\propto \tau^{N_x+1}+{\cal O}
\left(\tau^{N_x+2}\right), \end{array}
\end{align} 
\end{subequations}
as shown in  Tab.\ \ref{tab1} and in Fig.\ \ref{fig1a}. The data agree with 
the results of Ref.\ \onlinecite{quiro11} for the
overall error. For either $N_x=0$ or $N_z=0$ 
(UDD sequence) we find that $d$ scales as $\tau$, as expected for
a Hamiltonian with general decoherence. On the other hand, if $N_z=0$ for 
example and $H_{\text{qB}}$ is a pure dephasing Hamiltonian, i.e., no couplings 
with 
$\sigma_x^{(0)}$ or with $\sigma_y^{(0)}$ occur in the Hamiltonian, 
the norm distance scales as $d_{\text{UDD}}\propto \tau^{N_x+1}$.

Here we present the results of a given random configuration of the entries of 
$\hat J_0^{ij}$ and $\hat J_1^{i}$. We also checked that the scaling 
is the same for other random configurations.

\begin{table}[ht]
\begin{center}
\begin{tabular}{|c|c|c|c|c|c|c|c|}
\hline
 $N_z\setminus N_x$ & 0 & 1 & 2 & 3 & 4 & 5 & 6 \\
\hline
\hline
 0& 1.00 & 1.00 & 1.00  & 1.00 & 1.00 & 1.00 & 1.00 \\
\hline
 1& 1.00 & 2.01 & 2.00 & 2.01 & 2.01 & 2.01 & 2.00\\
\hline
 2& 1.00 & 2.01 & 2.97 & 3.01 & 3.01 & 3.01 & 3.02\\
\hline
 3& 1.00  & 2.00 & 2.99 & 4.01 & 4.07 & 4.09 & 4.09\\
\hline
 4& 1.00 & 2.00 & 3.00 & 4.01 & 4.99 & 4.98 & 5.05\\
\hline
 5& 1.00 & 1.95 & 3.02 & 3.95 & 5.01 & 5.98 & 6.05\\
\hline
 6& 1.00 & 1.98 & 3.01 & 3.94 & 5.00 & 5.84 & 6.95\\
\hline
\end{tabular}
\caption{\label{tab1} Low symmetry case: Scaling exponent $\zeta$ of the norm 
distance $ d(\tau)$ with $\tau$, the total duration of the sequence. The reported 
numbers are determined from the slope of the curve 
$d$ vs. $\tau$ in a double logarithmic plot.}
\end{center}
\end{table}

Different choices of the initial bath state $\rho_\text{B}$ of the form
\eqref{eq:rho_product}, i.e., varying the $|\gamma_i\rangle$,
 can affect the 
scaling of $d_x$, $d_y$ and of $d_z$, but not the overall scaling of 
$d$: If $ d_\gamma \propto \tau^{A_\gamma}$ then the leading order of the norm 
distance scales as $ d\propto \tau^{\min_\gamma\{A_\gamma\}}$. Hence the scaling exponent
reads $\zeta= \min_\gamma\{A_\gamma\}$. This is what the analytic
arguments require for QDD \cite{wang11,jiang11,kuo11}. Hence the analytic
bounds on the exponents are sharp for the low symmetry case.

\section{Case 2: High Symmetry}
\label{sec2}

We consider an SU(2) invariant isotropic central spin model 
with Heisenberg couplings.
We choose a Hamiltonian of the form of Eq.\ (\ref{eq:ham}) 
with $\hat{J}_{0}^{ij}=\alpha\lambda j_0^{ij}\mathbbm{1}$ and 
$\hat{J}_{1}^{i}=\lambda j_1^{i}\mathbbm{1}$, where $\alpha$ and $\lambda$ 
are two generic constants while $j_0^{ij}$ and $j_1^{i}$ are random numbers
between $-1$ and $1$. 

\begin{table}[ht]
\begin{center}
\begin{tabular}{|c|c|c|c|c|c|c|c|}
\hline
 $N_z\setminus N_x$ & 0 & 1 & 2 & 3 & 4 & 5 & 6 \\
\hline
\hline
 0& 1.99 & 2.00 & 2.00 & 2.00 & 2.00 & 2.00 & 2.00  \\
\hline
 1& 2.00 & 3.99 & 3.95 & 4.00 & 3.98 & 4.00 & 4.00  \\
\hline
 2& 2.00 & 4.00 & 5.99 & 6.13 & 6.00 & 5.99 & 5.99 \\
\hline
 3& 2.00 & 3.99 & 5.99 & 7.98 & 7.97 & 8.01 & 8.01\\
\hline
 4& 2.00 & 4.00 & 5.99 & 7.99 & 9.97 & 9.92 & 9.94\\
\hline
 5& 2.00 & 4.00 & 5.99 & 7.99 & 9.97 & 11.93 & 11.94 \\
\hline
 6& 2.00 & 4.00 & 6.00 & 7.97 & 9.95 & 12.01 & 13.95  \\
\hline
\end{tabular}
\caption{\label{tab2} High symmetry case: Scaling exponent $\zeta$ of the norm 
distance with $\tau$ for an isotropic Hamiltonian with general decoherence and for
 an initial density matrix such as in Eq.\ (\ref{eq:rho_0_1}). The scaling 
exponent are derived from a fit of the numerical curves for $d$ vs.\ $\tau$ in a 
double logarithmic plot.}
\end{center}
\end{table}

\begin{figure}[ht]
   \subfigure[]{ \includegraphics[width=0.5\columnwidth,clip]{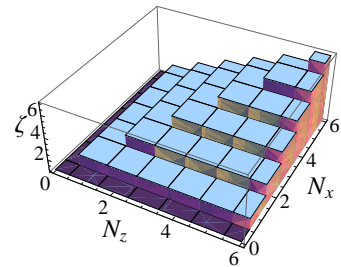} 
\label{fig1a}}
     \subfigure[]{\includegraphics[width=0.5\columnwidth,clip]{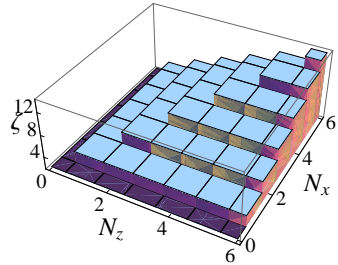} 
\label{fig1b}}
     \caption{(Color online) The data of Tabs.\ \ref{tab1} and \ref{tab2} are 
       graphically represented in panel (a) and in panel (b), respectively. 
       The numbers are rounded to their first digit. Here we have introduced 
       the notation here $d\propto \tau^\zeta$, where $\zeta$ stands for the 
       scaling exponents.  
       \label{fig1}
     }
\end{figure}

If the bath at $t=0$ is described by the following density matrix  
\begin{equation}
 \label{eq:rho_0_1}
\rho_\text{B}\propto \mathbbm{1}_\text{B},
\end{equation}  
the suppression of the decoherence is enhanced by a factor 2. 
The simulation for QDD yields the scaling exponents reported in 
Tab.\ \ref{tab2}. We deduce the following rules: 
for $N_x\neq 0$, $N_z\neq 0$ and 
\begin{subequations}
\begin{align}
 & \begin{array}{c}  N_x=N_z=N\rightarrow d\propto \tau^{2(N+1)}+{\cal O}
\left(\tau^{2N+3}\right), \end{array}
\\
 & \begin{array}{c}   N_x>N_z\rightarrow d\propto \tau^{2(N_z+1)}+{\cal O}
\left(\tau^{2N_z+3}\right), \end{array}
\\
 & \begin{array}{c}    N_x<N_z \rightarrow d\propto \tau^{2(N_x+1)}+{\cal O}
\left(\tau^{2N_x+3}\right), \end{array}
\end{align} 
\end{subequations}
If $N_x=0$ or $N_z=0$ the norm distance scales as $\tau^2$.
A graphical representation of the data is provided in Fig. \ref{fig1b}.

The state of Eq.\ (\ref{eq:rho_0_1}) is a completely disordered state where no 
particular spin direction or state is singled out. Such a state can be referred
 to as an ``infinite-temperature'' state. This is not unusual in NMR 
experiments where already at room temperature one finds 
$\hbar\omega_\text{L}/k_\text{B}T\approx 10^{-5}$, where $\omega_\text{L}$ is the 
Larmor frequency of a spin and $k_\text{B}$ is the Boltzmann constant.
This means that the thermal energy exceeds all internal energy scales by many 
orders of magnitude.

Note that for $\rho_\text{B}\propto \mathbbm 1_\text{B}$ the norm distance 
(\ref{eq:pF}) coincides with the partial Frobenius norm distance 
\cite{pasin11a,lidar08}.

\section{SU(2) invariance}
\label{sec3}

The appearance of the factor 2 can be explained in terms of the different 
parity of $H_\text{B}$ and $H_\text{qB}$ under spin rotations.
We write the Hamiltonian (\ref{eq:ham}) in the form
\begin{equation}
 \label{eq:H_Bp_Bm}
H=\mathbbm{1}_\text{S}\otimes A_0+\sum_{\mu\in\{x,y,z\}}\sigma_\mu \otimes A_\mu.
\end{equation} 
In Eq.\ (\ref{eq:H_Bp_Bm}) 
the operators $A_0$ and $A_\mu$ act only on the bath while $\mathbbm{1}_\text{S}$
 and $\sigma_\mu$ act only on the qubit.
Since the identity operator and the Pauli matrices form
a complete basis for all system operators the evolution operator
can be expanded according to
\begin{equation}
 \label{eq:U_Bp_Bm}
U(\tau,0)=\mathbbm{1}_\text{S}\otimes B_0(\tau)+\sum_{\mu=x,y,z}\sigma_\mu 
\otimes B_\mu(\tau),
\end{equation} 
where $B_0(\tau)$ and $B_\mu(\tau)$ are 
non-trivial functions of the operators $A_0$ and $A_\mu$ 
and of the switching functions $f_\mu(t)$,
see Eq.\ (\ref{eq:switching_functions}) in the Appendix
and Ref.\ \onlinecite{uhrig10b}.
For the sake of simplicity we will omit  the time dependence 
of the operators $B_0$ and $B_\mu$ from now on.
From the unitarity of $U(\tau,0)$ we conclude 
\begin{subequations}
  \label{eq:unitary_U_condit}
\begin{equation}
 \label{eq:unitary_U_condit1}
\mathbbm{1}_{\text{B}}=B_0B_0^\dagger+\sum_{\mu\in\{x,y,z\}}B_\mu B_\mu^\dagger
\end{equation} 
and
\begin{equation}
 \label{eq:unitary_U_condit2}
0=i\sum_{\mu,\nu\in\{x,y,z\}}\epsilon_{\mu\nu\kappa}B_\mu B_\nu^\dagger+
\left(B_0B_\kappa^\dagger+ \text{h.c.}\right)
\end{equation}
\end{subequations} 
for fixed $\kappa\in\{x,y,z\}$ and $\epsilon_{\mu\nu\kappa}$ being the 
Levi-Civita symbol.
In \eqref{eq:unitary_U_condit2} we omitted the non-singular
factor proportional to Pauli matrices because the vanishing
must be ensured by the bath operators.
In the Heisenberg picture the density matrix $\rho_0^{(\gamma)}$ (\ref{eq:rho_0})
 evolves according to
\begin{eqnarray}
 \label{eq:rho(T)}
\rho_0^{(\gamma)}(\tau)&=&U(\tau,0)\rho_0^{(\gamma)}U(\tau,0)^\dagger
\\
&=&U(\tau,0)\left(\rho_\text{S}^{(\gamma)}
\otimes\rho_\text{B}\right) U(\tau,0)^\dagger.
\end{eqnarray}
We trace out the bath and use the unitarity of $U(\tau,0)$ 
(\ref{eq:unitary_U_condit}) to obtain
\begin{equation}
 \label{eq:Tr_rhoT}
\text{Tr}_\text{B}\ \rho_0^{(\gamma)}(\tau)= T_1^{(\gamma)}+T_2^{(\gamma)}+ T_3^{(\gamma)}+ 
T_4^{(\gamma)}
\end{equation}
with
\begin{subequations}
\begin{align}
 \label{eq:T1}
T_1^{(\gamma)}:=&\rho_\text{S}^{(\gamma)}\text{Tr}_\text{B}\rho_\text{B}+
\sum_{\mu\in\{x,y,z\}}(c_{\mu,\mu}^{(\gamma)}-\rho_\text{S}^{(\gamma)})\ b_{\mu,\mu},
\\
 \label{eq:T2}
T_2^{(\gamma)}:= &\sum_{\mu,\nu\in \{x,y,z\}, \mu\neq \nu}
c_{\mu,\nu}^{(\gamma)}\ b_{\mu,\nu},
\\
 \label{eq:T3}
T_3^{(\gamma)}:= &\sum_{\mu\in\{x,y,z\}} d_\mu^{(\gamma)} b_\mu,
\\
 \label{eq:T4}
T_4^{(\gamma)}:= &-i\sum_{\mu,\nu,\kappa\in\{x,y,z\}}
\epsilon_{\mu\nu\kappa}\ \rho_\text{S}^{(\gamma)}\sigma_\kappa\ b_{\nu,\mu}.
\end{align}
\end{subequations}
The coefficients
\begin{subequations} 
\begin{eqnarray}
b_{\mu,\nu}&:=&\text{Tr}_{\text{B}}[B_\mu\rho_\text{B} B_\nu^\dagger]
\\
b_{\mu} &:=&\text{Tr}_\text{B}\left[B_0\rho_\text{B}B_\mu^\dagger\right]
\end{eqnarray}
\end{subequations} 
depend only on the bath operators while 
$c_{\mu,\nu}^{(\gamma)}:=\sigma_\mu\rho_\text{S}^{(\gamma)}\sigma_\nu$ and 
$d_{\mu}^{(\gamma)}:=\sigma_\mu\rho_\text{S}^{(\gamma)}-\rho_\text{S}^{(\gamma)}\sigma_\mu$ 
are functions  of the qubit operators only.
Note that for a pure dephasing model \cite{uhrig10b} the terms 
$T_2^{(\gamma)}$ and $T_4^{(\gamma)}$ do not appear.

We consider a global operator $\hat P_\nu$ that rotates all the spins of 
our system around the $\nu=x$, $y$, or $z$ axis by the angle $\pi$. Here we are
 interested in the
SU(2) invariant Hamiltonian such as  the one discussed in Sec.\ \ref{sec2}.
Then we have
\begin{subequations}
\begin{equation}
 \label{eq:B_even_odd}
\hat P_\nu B_0 \hat P_\nu^\dagger= B_0
\quad
\hat P_\nu B_\nu \hat P_\nu^\dagger=B_\nu\ ,
\end{equation}
\begin{equation}
\label{eq:B_even_odd_1} 
\hat P_\nu B_\mu \hat P_\nu^\dagger=-B_\mu\quad \text{for}\ \nu\neq\mu
\end{equation}
\end{subequations}
and therefore
\begin{subequations}
\begin{eqnarray}
 \label{eq:PbpmP}
b_\mu&=&\text{Tr}_\text{B}\left[B_0\rho_\text{B}B_\mu^\dagger\right]
\\
&=&\text{Tr}_\text{B}\left[\hat P_\nu B_0\hat P_\nu^\dagger\ 
\hat P_\nu\rho_\text{B}\hat P_\nu^\dagger\ \hat P_\nu B_\mu^\dagger
\hat P_\nu^\dagger\right]
\\
&=&-\text{Tr}_\text{B}\left[B_0 \ \hat P_\nu\rho_\text{B}\hat P_\nu^\dagger\ 
B_\mu^\dagger\right]
\label{eq:b_minus}
\end{eqnarray} 
\end{subequations}
for $\mu\neq \nu$.
Thus, if $\rho_\text{B}$ is invariant under rotation $\hat P_\nu$, 
which is the case for $\rho_\text{B} \propto \mathbbm 1_\text{B}$, 
we can conclude from  \eqref{eq:b_minus} that 
$b_\mu=0$ for $\mu\in\{x, y, z\}$. In fact,
the analogous argument also implies $b_{\mu,\nu}=0$ for
$\mu\neq \nu$, though we will
not use this fact here.  The  condition $\mu\neq \nu$ is needed to 
ensure that we can flip the sign of the two factor
$B_\mu$ and $B_\nu^\dag$ separately.

If the coefficients $b_\mu$ vanish only the terms proportional to $b_{\mu,\nu}$ 
remain in Eq.\ (\ref{eq:Tr_rhoT}). We know from the analytic
properties of the QDD sequence that the operators $B_\mu$ with
$\mu\in\{ x,y,z\}$ all scale at least with $\tau^{N_\textrm{min}+1}$ where 
$N_\textrm{min}:=\min(N_x,N_z)$ \cite{wang11,jiang11,kuo11}
supported by numerical results in Ref.\ \onlinecite{quiro11} and in
the present work.
Hence the coefficients $b_{\mu}$ scale with $\tau^{N_\textrm{min}+1}$
and the coefficients $b_{\mu,\nu}$ with $\tau^{2N_\textrm{min}+2}$.
Hence the vanishing of the $b_\mu$ terms in Eq.\ (\ref{eq:Tr_rhoT})
automatically reduce the decoherence by doubling
the exponent in the scaling 
$d\propto \tau^{N_\textrm{min}+1} \to \tau^{2N_\textrm{min}+2}$
with the total duration $\tau$ of the sequence.

Note that for a model of pure dephasing, e.g., only
$\sigma_z^{(0)}$ appears in \eqref{eq:ham_b}, we do not
need the symmetry with respect to two operators $\hat P_\mu$.
It is sufficient to have \emph{either} $\hat P_x$ or $\hat P_y$ which
invert $\sigma_z$ so that we can conclude that
$b_z$ vanishes in order to know that the scaling exponent doubles.
This was already seen in the numerical data presented and analyzed in
Ref.\ \onlinecite{pasin11a}.

\section{Case 3: Mixed Symmetry}
\label{sec4}
\begin{table}[h]
\begin{center}
\subtable[]{
\begin{tabular}{|c|c|c|c|c|c|c|c|}
\hline
 $N_z\setminus N_x$ & 0 & 1 & 2 & 3 & 4 & 5 & 6 \\
\hline
\hline
 0& 0.99 & 1.00 & 1.00 & 1.00 & 1.00 & 1.00 & 1.00  \\
\hline
 1& 1.00 & 2.00 & 2.00 & 2.00 & 2.00 & 2.00 & 2.00  \\
\hline
 2& 1.00 & 1.99 & 3.01 & 3.00 & 3.00 & 3.00 & 3.00 \\
\hline
 3& 1.00 & 2.00 & 3.00 & 4.00 & 4.00 & 4.00 & 4.00\\
\hline
 4& 1.00 & 2.01 & 3.00 & 4.02 & 5.00 & 5.00 & 5.01\\
\hline
 5& 1.00 & 2.00 & 3.00 & 4.00 & 5.02 & 6.00 & 6.00 \\
\hline
 6& 1.00 & 2.00 & 3.00 & 3.98 & 5.00 & 6.00 & 6.99  \\
\hline
\end{tabular}
\label{tab3a}}
\subtable[]{
\begin{tabular}{|c|c|c|c|c|c|c|c|}
\hline
 $N_z\setminus N_x$ & 0 & 1 & 2 & 3 & 4 & 5 & 6 \\
\hline
\hline
 0& 1.98 & 2.00 & 2.00 & 2.00 & 2.00 & 2.00 & 2.00  \\
\hline
 1& 2.00 & 2.00 & 2.00 & 2.00 & 2.00 & 2.00 & 2.00  \\
\hline
 2& 2.00 & 3.99 & 6.00 & 5.75 & 5.99 & 5.99 & 5.99 \\
\hline
 3& 2.00 & 4.00 & 5.80 & 4.00 & 4.00 & 4.00 & 4.00\\
\hline
 4& 2.00 & 4.00 & 6.00 & 7.96 & 9.96 & 10.27 & 10.30\\
\hline
 5& 2.00 & 3.97 & 5.97 & 7.97 & 9.85 & 6.00 & 6.00 \\
\hline
 6& 2.00 & 4.00 & 5.99 & 7.99 & 9.95 & 11.93 & 13.92  \\
\hline
\end{tabular}
\label{tab3b}}
\caption{\label{tab3} 
  Mixed cases: In Table (a) the Hamiltonian is SU(2) invariant
  while the bath is initially prepared in a product state. In Table (b) the 
  Hamiltonian is asymmetric of the form in Eq.\ (\ref{eq:ham}), the entries of 
  $\hat J_0^{ij}$ and $\hat J_1^{i}$ are all randomly chosen, and 
  $\rho_\text{B}\propto \mathbbm{1}_\text{B}$.}
\end{center}
\end{table}

Here we analyze the off-diagonal cases of Tab.\ \ref{tab0}. They are 
characterized either by a SU(2) invariant Hamiltonian and a low-symmetry bath 
state or by a low-symmetry Hamiltonian with a high-symmetry initial bath
state $\propto \mathbbm{1}_\text{B}$.
Because the Hamiltonian and the density matrix have 
a different degree of symmetry the analytical argument of
 Eq.\ (\ref{eq:PbpmP}) for $b_\mu$ does not hold anymore. 
\begin{figure}[ht]
       \includegraphics[width=0.5\columnwidth,clip]{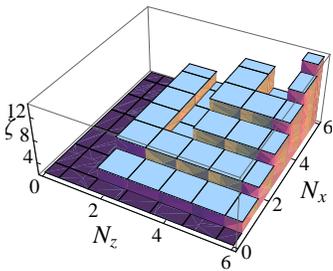} 
     \caption{(Color online) The data of 
       \ref{tab3} (b) is graphically represented. 
       The numbers are rounded to their first digit. 
       The notation $\zeta$ refers to the scaling exponents of the norm 
       distance $d\propto \tau^\zeta$.  
       \label{fig2}
     }
\end{figure}

The numerically found scaling exponents are reported in Tabs.\ \ref{tab3} and 
depicted in Fig. \ref{fig2}. The coupling constants used to derive the scaling 
exponents of this Table are the same as those used for the simulation of the 
SU(2) invariant Hamiltonian or the ones used
for the asymmetric Hamiltonians used in
Secs.\ \ref{sec1} and \ref{sec2}.

If the Hamiltonian is SU(2) symmetric and the bath is initially prepared in a product state
the scaling exponents look the same as those of Tab. \ref{tab1}.
If the Hamiltonian is asymmetric and 
$\rho_\text{B}\propto \mathbbm{1}_\text{B}$ we find for $N_x=N_z=N$ that 
the scaling  exponent is $N+1$ for $N$ odd and $2(N+1)$ for $N$ even. 
For $N_x>N_z$ the 
exponent is either $N_z+1$ if $N_z$ is odd or $2(N_z+1)$ if $N_z$ is even while 
for $N_x<N_z$ we find that $d$ scales as $2(N_x+1)$. We cannot provide any explanations for this alternating 
behavior of the scaling exponents and for the reason why, for $N_x>N_z$, it depends only on the number of pulses of the 
inner sequence. This is still an open question for future investigation.

Furthermore, it is worth mentioning that peculiar choices
of the couplings imply non-generic behavior.
If the couplings $j_1^i$ is the same for all $i$ the QDD sequence behaves 
like a UDD sequence with  $N_z$ pulses for $N_x$ odd, as if the qubit were subject only 
to pure dephasing: Either $A_x\neq 0$ or  $A_y\neq 0$ in Eq.\ \ref{eq:H_Bp_Bm}. 
It is not clear why the QDD behaves like a UDD sequence in this case. 
A possible explanation is that the 
product state we take as initial state of the bath is not the most general one. 
It is not entangled. In the Appendix we analyze the first 
three cumulants of the evolution operator for the case $N_x=1$. We find that 
they contain only one qubit operator, either $\sigma_x$ or $\sigma_y$, and that
 the dephasing term proportional to $\sigma_z$ is always zero.  For $N_x$ even 
we recover the results of Table \ref{tab1}.

\section{Conclusions}
\label{conclusions} 

We investigated the influence of the state
of the bath in suppressing general decoherence
by means of a QDD sequence. The performance of the sequence is measured by the 
norm distance $d$ which is essentially the norm of the remaining 
decoherence. Thus the performance
is quantified by the scaling of $d$ with $\tau$, the total 
duration of the sequence. Recent papers \cite{wang11,jiang11,quiro11,kuo11} 
proved the properties of QDD and clarified the dependence of the 
scaling on the number of pulses $N_x$ and $N_z$: The overall scaling of QDD 
is given by $\zeta:=\min\{N_x,N_z\}+1$ independent of the details of the 
environment, i.e., $d\propto \tau^\zeta$.
In this sense QDD is a universal sequence for 
general decoherence such as UDD is a universal sequence for
pure dephasing.

In the present work, we have shown that the actual performance
of QDD can be even better than expected on the basis of the general mathematical
arguments. This improvement occurs if the Hamiltonian and the bath state
are highly symmetric, for instance, if the Hamiltonian is
spin isotropic and the bath is prepared initially 
in a completely disordered state. Then we found both numerically and
analytically that the exponent of the scaling with $\tau$ acquires an 
additional factor of 2: $d\propto \tau^{2\zeta}$.
The same was already observed for the UDD sequence applied
to pure dephasing in Ref.\ \onlinecite{pasin11a}.

We emphasize that this result is by no means at odds with the
proofs of universality \cite{wang11,jiang11,kuo11}.
The general proofs refer to the worst case for decoherence.
They determine whether a certain operator (Pauli matrix) 
for the qubit occurs or not \emph{irrespective} of the bath operator
to which it is multiplied. The underlying idea is that for any
non-vanishing bath operator there is a bath state such that
the qubit state is influenced in a non-trivial way. 
Hence decoherence occurs.

But for certain choices of the bath state even a non-vanishing
bath operator may have a vanishing effect on the quantum bit
if its partial bath trace vanishes.
Then no decoherence is induced by this particular term.
This is the effect which enhances the performance of the QDD sequence
for highly symmetric situations. We summarize that $\min\{N_x,N_z\}+1$ is a 
lower bound for $\zeta$.

We stress that the found phenomenon is relevant for realistic situations.
Complete or partial spin symmetry in the Hamiltonian is a standard feature.
A completely disordered bath state is also an excellent starting point
in the description of baths of nuclear spins. Their mutual interaction
is so small in energy that even room temperature suffices
to disorder the nuclear spins completely.

In general, we  conclude that the more asymmetric the bath Hamiltonian, its
coupling to the qubit, and in particular its initial state are, 
the lower the exponent $\zeta$ is of the leading non-vanishing
power in the total duration $\tau$ of the sequence inducing decoherence.
The same is true of UDD sequences for pure dephasing.
So far, we focussed on spin baths which allow for completely disordered,
infinite temperature states. It is an interesting question for future research
whether a similar phenomenon can occur in other baths such as bosonic
ones.

Experimental research is also called for. To our 
knowledge, there exist studies on the influence of the initial state,
see for instance Ref.\ \onlinecite{alvar10c}, but they
focus on the initial state of the system. Discussions of the influence
of the initial state of the bath, which was our focus here,
are scarce \cite{li08a}.
Moreover, it must be distinguished between studies of iterated
cycles of sequences with exponential decay rates \cite{ajoy11} and studies
of a single sequence displaying decoherence with a particular
power law 
\cite{uhrig07,lee08a,uhrig08,yang08,west10,wang11,jiang11,kuo11}.

Of course, it is difficult to measure the exponents directly.
But we suggest to demonstrate experimentally that the performance of a 
QDD or a UDD sequence is lowered if the symmetry of either the Hamiltonian 
or of the initial bath state is lowered. This would already 
be a smoking gun evidence for the essence of the present theoretical finding.

\begin{acknowledgments}
We would like to thank Gregory Quiroz and Daniel A. Lidar for helpful 
discussions. The financial support by the grant  UH 90/5-1
of the DFG is gratefully acknowledged.
\end{acknowledgments}

\appendix
\section{Magnus expansion of the evolution operator}

In general, it is complicated to find a universal argument that explains the 
scaling of QDD if the symmetry of the system is not well defined. The reason why 
some exponents deviate from the analytically predicted 
value depends on the possible vanishing
of some terms in the Magnus expansion \cite{haebe76,magnu54} of the evolution 
operator. The form of such terms strongly depends on the form of the 
Hamiltonian and of the symmetry of the bath state. 

If no general conclusions can be drawn on the parity of these 
terms and of the density matrix under a rotation $\hat P_\mu$, one
way to proceed is to calculate the cumulants of the expansion 
explicitly and to analyze which terms determine the power law of $d$ with $\tau$. 
Here we provide an analysis of the first cumulants of the Magnus expansion  
for $N_x$ odd for the data of Tab.\ \ref{tab3}(a). 

For a generic instant $t\in[0,\tau]$ we write the Hamiltonian (\ref{eq:ham}) as 
\begin{eqnarray}
 \label{eq:H(t)A}
H(t)=
H_\text{B}+\sum_{\mu=x,y,z}\sigma_\mu f_\mu(t) A_\mu.
\end{eqnarray}
The switching functions \footnote{The form of the switching function can be 
easily understood if one remembers that an $X$ $\pi$-pulse changes the signs in 
front of the $Y$ and of the $Z$ coupling, while a $Z$ $\pi$-pulses 
changes the signs in front of the $X$ and of the $Y$ coupling.}
 $f_\mu(t)$ are the effect of the stroboscopic pulsing of the qubit. 
They are piecewise constant functions with values $\pm 1$
\begin{subequations}
\label{eq:switching_functions}
\begin{equation}
 \label{eq:fz}
 f_z(t)=(-1)^j \quad  \text{for}\quad t\in (t^x_j,t^x_{j+1}],
\end{equation}
\begin{equation}
 \label{eq:fy}
 f_y(t)=(-1)^k \quad  \text{for}\quad t\in (t^z_{j,k},t^x_{j+1,k}],
\end{equation}with $f_y(0)=1$ and $f_y(\tau)=-1$, and
\begin{equation}
 \label{eq:fx}
 f_x(t)=f_z(t) f_y(t).
\end{equation}
\end{subequations}

The evolution operator can be written in terms of the cumulants $\bar{H}^{(n)}$ as
\begin{eqnarray}
 \label{eq:U_cumulants}
U(\tau,0)&=&\exp\{-i\tau\sum_{n=1}^\infty \bar{H}^{(n)}\}
\end{eqnarray} with $\tau\bar{H}^{(n)}\propto \tau^n$.
The first and the second cumulants are defined \cite{haebe76} as $\tau \bar{H}^{(1)}=\int_0^\tau H(t) \text{d}t$, 
$\tau \bar{H}^{(2)}=-\frac{i}{2 \tau}\int_0^\tau \text{d}t_1\int_0^{t_1}\text{d}t_2 \left[H(t_1),H(t_2)\right]$.
From Eqs. (\ref{eq:switching_functions}) it is straightforward to verify that $\int_0^\tau \text{d}t f_\mu(t)=0$ for $\mu=x$, $y$ or $z$ and for $N_z=N_x=1$. The first cumulant is proportional to the bath Hamiltonian
\begin{equation}
 \label{eq:H1} \tau \bar{H}^{(1)}=\tau H_\text{B}.
\end{equation} 
For the second cumulant one finds
\begin{eqnarray}
 \label{eq:H2}  2 i \tau \bar{H}^{(2)}= \sum_{\mu=x,y,z}\sigma_\mu I_2^\mu\left[H_\text{B},A_\mu\right]\\
 + \sum_{\mu,\nu=x,y,z}I_2^{\mu,\nu}
\left[\sigma_\mu A_\mu,\sigma_\nu A_\nu\right],
\end{eqnarray}
with the integrals 
\begin{subequations}
 \label{eq:integrals}
\begin{equation}
 \label{eq:I2mu}
I_2^\mu:=\int_0^\tau \text{d}t_1\int_0^{t_1}\text{d}t_2\left(f_\mu(t_2)-f_\mu(t_1)\right)
\end{equation}
and
\begin{equation}
 \label{eq:I2munu}
I_2^{\mu,\nu}:=\int_0^\tau \text{d}t_1\int_0^{t_1}\text{d}t_2 f_\mu(t_2)f_\nu(t_1).
\end{equation}
\end{subequations}
The integrals (\ref{eq:I2mu}) and (\ref{eq:I2munu}) can be easily evaluated, we report below only those that are different from zero. For $N_z=N_x=1$ one finds
\begin{equation}
 \label{eq:eval_I2}
I_2^y=\frac{\tau^2}{4},\quad I_2^z=\frac{\tau^2}{2},\quad  I_2^{x,z}=-I_2^{z,x}=
-\frac{\tau^2}{4}.
\end{equation}
The second cumulant then becomes
\begin{align}
 \label{eq:H2_fin}  2 i \tau \bar{H}^{(2)}&= \frac{\tau^2}{4}\sigma_y 
\left[H_\text{B},A_y\right]+\frac{\tau^2}{2}\sigma_z \left[H_\text{B},A_z\right]
\nonumber\\
&+i\frac{\tau^2}{2}\sigma_y\left[ A_x,A_z\right]_+,
\end{align}where the notation $[\ ,\ ]_+$ stands for an anticommutator. 
It is interesting to notice that $\bar{H}^{(2)}$ does not contain any terms 
proportional to $\sigma_x$.

If the Hamiltonian is SU(2) invariant and the coupling constants $\hat{J}_0^{ij}$ and
$\hat{J}_1^{i}$ are equal and independent of the indexes $i$ and $j$ 
($\hat{J}_0^{ij}= \alpha\lambda$ and $\hat{J}_1^{i}= \lambda$)
the commutators in Eq.\ (\ref{eq:H2_fin}) vanish because the Pauli matrices 
anticommute. This is true for a central spin model.

On the other hand, if 
$A_\mu\equiv \sigma_\mu^{(1)}$ (e.g. for a spin chain) the commutators in 
Eq.\ (\ref{eq:H2_fin}) are different from zero.
The anticommutator becomes $\left[A_x,A_z\right]_+
=2i\sum_{i,j}(\hat J_1)_{xx}(\hat J_1)_{zz}(1-\delta_{ij})\sigma_x^{(i)}
\sigma_z^{(j)}$
which implies that it vanishes for a spin chain because 
the qubit is coupled to a single site only, i.e., $i=j$ holds.

As usual we are interested in the difference between the evolved density 
matrix $\rho_0^{(\gamma)}(\tau)=U(\tau,0)\rho_0^{(\gamma)} U(\tau,0)^\dagger$ and the 
initial one. We find 
\begin{align}
 \label{eq:diffrho_isotropic}
\nonumber
 &\text{Tr}_\text{B}\left\{\rho_0^{(\gamma)}(\tau)-\rho_0^{(\gamma)}\right\}
=\frac{\tau^2}{4}[\sigma_y,\rho_\text{S}^{(\gamma)}]\ \text{Tr}_\text{B}
\left\{\rho_\text{B}\left[ A_x,A_z\right]_+\right\}
\\
& \qquad
+\frac{\tau^4}{16}\ c_{yy}^{(\gamma)}\ \text{Tr}_\text{B}\left\{\left[ A_x,A_z\right]_+
\rho_\text{B}\left[ A_x,A_z\right]_+\right\}
\end{align} 
where the hermitecity $A_\mu^\dagger=A_\mu$ was used. In writing 
Eq.\ (\ref{eq:diffrho_isotropic}) we neglected the contributions coming 
from the first cumulant because they do not alter the qubit-operator content 
of the norm distance.

If  $\rho_\text{B}\propto \mathbbm{1}_\text{B}$ the terms with the trace over the 
bath vanish for a SU(2) invariant Hamiltonian (see Sect.\ \ref{sec2}) 
while they are finite for an asymmetric model such as the one in  
Eq.\ (\ref{eq:ham}) in Sect.\ \ref{sec1}.

In order to understand why the distance norm scales with
exponents $N_z+1$ for $N_x=1$ 
(and in general for $N_x$ odd) in Tab.\ \ref{tab3}(a), some knowledge on the 
third cumulant $\bar H^{(3)}$ is required. This cumulant is defined 
\cite{haebe76} as
\begin{align}
\label{eq:H3_definition}
\nonumber 
&-6\tau \bar H^{(3)}=\int_0^\tau \text{d}t_1\int_0^{t_1}\text{d}t_2\int_0^{t_2}\text{d}t_3\ 
\times
\\ 
&\left\{[H(t_3),[H(t_2),H(t_1)]] + [H(t_1),[H(t_2),H(t_3)]]\right\}.
\end{align}
The commutators $[H(t_2),H(t_1)]$ and $[H(t_2),H(t_3)]$ have the same operator 
content as $\bar H^{(2)}$, but  differ in their prefactors and in their
time dependence. One can verify that $[H_\text{B},\bar H^{(2)}]=0$ if the bath 
Hamiltonian is SU(2) invariant due to the anticommutation of the Pauli 
matrices. For the same reason one finds that 
\begin{equation}
\label{eq:H3}
 \left[\sum_\mu \sigma_\mu A_\mu, \bar H^{(2)}\right]\propto\sum_{\mu,\nu}
\epsilon_{\mu,y,\nu}\ \sigma_\nu\otimes\sigma_x^{(i)}\sigma_y^{(j)}\sigma_\mu^{(k)},
\end{equation} where $i\neq j\neq k$. 
Equation (\ref{eq:H3}) only provides the operator contents of Eq.\ 
(\ref{eq:H3_definition}). In order to eliminate the time dependence one must 
substitute Eq.\ (\ref{eq:H3}) into Eq.\ (\ref{eq:H3_definition}) and integrate 
over $t_1$, $t_2$ and $t_3$. Each operator $A_\mu$ brings a switching function 
$f_\mu(t)$ with it, such that the integration in (\ref{eq:H3_definition}) 
yields the coefficients 
\begin{equation}
 \label{eq:I3} I_3^{\alpha,\beta,\gamma}=\int_0^\tau \text{d}t_1\int_0^{t_1}\text{d}t_2
\int_0^{t_2}\text{d}t_3f_\alpha(t_1)f_\beta(t_2)f_\gamma(t_3).
\end{equation}
The indices $\alpha$, $\beta$ and $\gamma$ can be equal to $x$, $y$, and 
$z$. We have checked numerically 
that for $N_x=1$ and $N_z=2$ the only non-zero contributions are given by 
$I_3^{x,z,z}$, $I_3^{z,x,z}$ and $I_3^{z,z,x}$ corresponding to the qubit operator 
$\sigma_x$. Thus the third cumulant $\bar H^{(3)}$ is a term of pure dephasing 
that can be suppressed by means of a $Z$ sequence of $\pi$ pulses. 
From the numerical results we expect that the same argument holds in general 
for higher cumulants and for $N_x$ odd.
 
We also checked our results for a spin chain. We find the same results as for 
a central spin model in the cases of low symmetry, high symmetry and in the 
case of an asymmetric Hamiltonian with $\rho_\text{B}\propto 
\mathbbm 1_\text{B}$. 
Discrepancies are found for the SU(2) invariant Hamiltonian in combination 
with the product state $|\psi_\text{B}\rangle$ (\ref{eq:rho_product}). A 
possible explanation is provided by the commutators in Eq.\ (\ref{eq:H2_fin}) 
that do not vanish for a spin chain with $A_\mu=\sigma_\mu^{(1)}$. Hence the 
precise topology of the model matters for the case of mixed degree of symmetry.

\bibliographystyle{apsrev}
\bibliography{liter10}

\end{document}